\documentclass[preprint,12pt]{elsarticle}

\usepackage{natbib}
\usepackage{epsfig}
\usepackage{amssymb}

\usepackage{rotating}
\newcommand{\bro}{\begin{rotate}}
\newcommand{\ero}{\end{rotate}}

\newcommand{\bc}{\begin{center}}
\newcommand{\ec}{\end{center}}
\newcommand{\bs}{\bigskip}

\newcommand{\ii}{\'{\i}}
\newcommand{\az}{\'a}
\newcommand{\ez}{\'e}
\newcommand{\oz}{\'o}

\newcommand{\bt}{\begin{tabular}}
\newcommand{\et}{\end{tabular}}
\newcommand{\bta}{\begin{table}}
\newcommand{\ete}{\end{table}}
\newcommand{\mtc}{\multicolumn}

\newcommand{\bp}{\begin{picture}}
\newcommand{\ep}{\end{picture}}
\newcommand{\bfi}{\begin{figure}}
\newcommand{\efi}{\end{figure}}
\newcommand{\dfrac}{\displaystyle\frac}

\begin{document}
\begin{frontmatter}

\title{Non extensive statistic of Tsallis in the heartbeat of healthy humans.} 

\author[ritto]{P. A. Ritto}
\ead{augustomijangos@yahoo.com}
\address[ritto]{P.O. Box 15, Admon. Centro 97001, M\'erida, M\'exico.}

\begin{abstract}
It is studied the MIT-BIH Normal Sinus Rhythm Database using a statistical technique of analysis, that is based on the Wavelet and Hilbert Transforms. With that technique, it was previously found, that there is a collective and intrinsic dynamical behavior up to a scale of 64 heartbeats. Now it is shown, that using the Biorthogonal wavelet bior3.1 such a behavior reaches the scale 1024. That result confirms, that the circulatory system is out of equilibrium. According to the Statistical Mechanics of Tsallis, and a recent interpretation of G. Wilk {\it et al.} respect to the non extensive parameter $q$, the healthy human being is characterized by $q=1.7\pm 0.02$.
\end{abstract}


\end{frontmatter}

\section{Introduction.}

The traditional way of measuring the cardiac activity is through the electrocardiogram (ECG), that records the voltage generated at the heart. The study of the ECG is quite complex because it depends on the position of the electrodes, the electrical properties of the skin, and the physiological state of the human being~\cite{position}. There are several techniques traditionally used in order to analyze the heartbeating. Those techniques are so simple as the ones measuring the cardiac mean frequency or as complex as those focused on the physical interpretation of the ECG signal. The main purpose of those techniques used in the study of the ECG is
 to find the physical and mathematical properties to be applied for the prognostics of the sickness of the heart. In order to achieve those objectives, it is common the study of the time series corresponding to an ECG which elements are the voltages generated at the heart. An ECG is obtained sticking electrodes to the skin, amplifying the voltage, and then sending the data to a storing place. The right sampling frequency satisfies physiological and technical standards~\cite{aha}, that are applied depending of the kind of the electrical signal, the amount of storing memory, the level of noise of an ECG signal, and of the scale of study of those data. During the years of study of the ECG it has been found, that applying a well known topological theorem of Takens-Sauers (T-S)~\cite{takens} it is feasible to analyze the data of an ECG selecting only the maximum --named ``R''-- of each beat. Taking each maximum, it is feasible to rearrange the time series into an equivalent one, as a consequence, it is possible to do a faster analysis. Through the years it has been found that, that kind of analysis is enoughly robust to be used as a predicting and prognosticating tool for several cardiac diseases.

There are several statistical techniques that use the T-S theorem in order to discover characteristics that are common to a set of patients~\cite{robustas,nature}. One of them, based on the Wavelet and Hilbert transforms is of our interest, because unveils a hidden behavior of the heart, that is reminiscent of a system out of equilibrium. Such a technique of analysis follows a procedure that is similar to the one used in the study of some nonlinear mechanical systems~\cite{meca}, that is useful to obtain unusual information of the cardiac dynamics~\cite{nature}-\cite{gamaost}. The name of that technique is Cumulative Variation of Amplitude Analysis (CVAA), and follows the steps: (i) Select a wavelet~\cite{mathworks}: $\Psi^{(m)}$ ($m$ is the $m$-{\it th} moment, that indicates that $\int_{-\infty}^{+\infty}x^{m}\cdot\Psi^{(m)}(x)\,dx=0$) 
in order to analyze the data of an ECG: $x_i\equiv x(t_i),i=1,2,...,n$, (ii) Select a set of temporal scales: $s\equiv 1,2,...,s_o$, where $n\in\aleph$, (iii) Apply a Continuous Wavelet Transform to the time series~\cite{slp,mathworks}: $W_i = \sum_j \Psi^{(m)}\left[s_j^{-1}(t_i-t_j)\right]\cdot x_i$, where $W_j$ are the coefficients of the Wavelet Transform, (iv) Apply the Hilbert Transform to the time series: $h_i\equiv W_i+\imath H_i$, where $H_i\equiv H(W_i)$~\cite{nature,mathworks}, (v) To obtain the amplitudes $A_i\equiv\sqrt{W_i^2 + H_i^2}$ for the elements of the new time series, (vi) Normalize the area under the distribution of amplitudes, and (vii) Rescale that distribution such as $A_i\rightarrow A_i\cdot P_{\rm max}$ and $P(A_i)\rightarrow P(A_i)/P_{\rm max}$, where $P_{\rm max}$ is the maximum of the normalized distribution. A decade ago, the CVAA was successfully applied to a set of time series corresponding to healthy people, giving as a result of that study a common Gamma distribution: 
$G_{\nu}(x)=b^{\nu +1}x^\nu e^{-bx}/\Gamma(\nu+1)$, where $b\equiv \nu/x_o$, being $x_o$ the value that maximizes such distribution. The wavelet used in that study is the Gaussian: $g_n(x)=C_n\dfrac{d^n}{dx^n}e^{-x^2}$. The characteristic parameter of the Gamma distribution is $\nu=1.8\pm0.1$ during the day (6 h.), while during the night (6 h.) it is $\nu=1.4\pm0.1$. The fact of a common Gamma distribution suggests, that there is an intrinsic dynamics at the heart characteristic of a system out of equilibrium~\cite{nature}. Even more, that collective behavior is typical of system that show a phase transition~\cite{renorma}. It is not still clear, the difference between the parameters during the day and during the night hours; according to~\cite{nature} that change is due to the circadian cycle that affects the nervous system. In this work, with the aim of exploring the behavior of the data according to the wavelet chosen, it is studied by the long term the same set of time series corresponding to healthy human beings. In order to study the effect of each wavelet on the result of CVAA, that one selected~\cite{mathworks} correspond to three quite different families: An Orthogonal (Daubechies), Biorthogonal, and Non Orthogonal: The Gaussian function, which is chosen by the authors of CVAA. In this study, it 
is also identified, the temporal scales for the uniqueness of the Gamma function. 

\section{Analysis of the data.}

The CVAA is used on the data base Physionet: MIT-BIH Normal Sinus Rhythm Database (nsrdb)~\cite{physbank}, the same that has been used by the authors of this technique. It is selected a set of 18 patients: 5 males of 26 to 45 years old, and 13 females of ages between 20 and 50. The time series are obtained directly of that data base at Physionet of ECGs, recorded at 128 samples per second. The period of each ECG is of the order of 24 h. The interbeat intervals RR are obtained using a software provided at Physionet. The length of each time series is of the size of $10^5$ heartbeats. It is applied the CVAA to the data using the next wavelets~\footnote{The notation for wavelets is used according to Ref.~\cite{mathworks}. For the Orthogonal and Non Orthogonal wavelets it is used : wavabrN, where small letters indicate the abbreviation for the wavelet``wavabr'' and ${\rm N}$ corresponds to the ${\rm N-1}$-th moment. For the Biorthogonal wavelet the notation is: bior${\rm N_r.N_d}$, where ${\rm N_r.N_d}$ corresponds to the moments ${\rm N_r}$ of reconstruction and ${\rm N_d}$ of decomposition.}: Daubechies: db1, db2, db3, Biorthogonal: bior3.1, bior3.3, bior3.5, and Gaussian: gaus1, gaus2, gaus3, following the steps suggested at the Section~1. In this study, the moments of the wavelets are not higher because the patients at the hospital are supposed to be almost in rest during the day, and that they don't use to follow activities that modify their physiological state in a specific way: At least that hasn't been indicated at the data base Physionet. The range of selected scales for our study is 64 ($\approx 1{\,\rm min}$)$-1024$ ($\approx 10{\,\rm min}$) 
heartbeats. The goal is to explore the behavior at scales higher than 64~\cite{nature}, in order to find the limit of fractality of the time series of heartbeats, using CVAA and the set of wavelets and scales. The long term study of the heartbeating of healthy people starts applying a filter to the time series of RR intervals, according to the algorithm at~\cite{filtra}, which avoids the spurious data. After that step, it is followed the procedure of CVAA for each of the wavelets and scales already selected. The complete analysis of the cardiac time series is done using a software for the analysis of ECGs that has been previously described at Ref.~\cite{acalan}. 

\section{Results and discussion.}

In figure~1, it is shown an instance of data collapsing of the cardiac data, which has been obtained after applying the CVAA. The analysis corresponds to the wavelet bior3.1 and the scale 1024. The parameters of the fit are $\nu=1.46\pm 0.03$ and $\chi^2=0.26$~\footnote{In this document $\chi^2\equiv\chi^2/DOF$.}. In the tables~1-4, it is presented each one of the fitting parameters for a wavelet and scale of study. It is possible to see at those tables, that as the scale of analysis increases the fitting parameters vary up to $~80\%$. The fractality, which is understood as the statistical invariance after rescaling, is only seen using the wavelet bior3.1, and for it corresponds an almost constant fitting parameter $\nu=1.43\pm 0.03$ for the scales 64 to 1024. As it is seen at figure~1 and at the table~1, the wavelet bior3.1 provides a good data collapse, and numerical fitting for the cardiac data. The analyses corresponding to other wavelets show that the fitting parameters change mainly inside those scales. Additionally, for the wavelets different than the bior3.1, the data collapse is missed starting at the scales: $[8,6,5]\times 2^6$, $[13,7]\times 2^6$, $[3,3,5]\times 2^6$ corresponding to [db1,db2,db3], [bior3.2,bior3.5] and [gaus1,gaus2,gaus3] respectively. It is important to comment that if the data are not filtered according to Ref.~\cite{filtra}, the distributions behave in a quite different way, not necessarily as a Gamma function. In other words, the data don't collapse into a single distribution; that result is independent of the wavelet selected.

The results obtained during this work show that, it is not necessary to cut the time series in shorter pieces {\it e.g.} 6 h. --as it is done at Ref.~\cite{nature}-- in order to reduce the non stationarities, and to obtain better fits to the parameters of the Gamma function. Periods of 6 h. during the day and 6 h. during the night doesn't show meaningful difference in $\nu$ if it is used the wavelet bior3.1 for CVAA. On the other hand, for a period of the order of 1 h., it is feasible to obtain acceptable data collapse but without using the Hilbert Transform~\cite{japan}. That suggests that the difference of $\nu$ during day and night hours is due to the wavelet selected, the size of the time series, and not necessarily to the non stationary phenomena at an ECG.

The integral of the Gaussian wavelet, $I=\int_{-\infty}^{+\infty}G_\nu^{(m)}(x)dx$ , isn't null as corresponds by definition, but this fact apparently doesn't affect a CVAA at short scales~\cite{nature}. According to the results shown here, $I\approx 0$ is important for changing the value of the fitting parameters --mainly $\nu$--, for the scales shown at tables~1-4. Ref.~\cite{physa,agili} suggests that the physiological state of a patient is not supposed to show a change at the distributions after the normalization. However, one of the results shown at Ref.~\cite{nature} seems to contradict that sentence. Here, the selection of the wavelet used in CVAA is fundamental at the time of studying long periods of an ECG. The wavelet Biorthogonal, moments 3,5 and the Orthogonal Daubechies, moments 0-2 let to obtain a good data collapse, and an acceptable fit to the data for scales higher than 64. Those graphical
 and numerical results are missed slower for the Biorthogonal and Orthogonal wavelets, than for the Gaussian one, as the scale of analysis increases. The difference of irregularity of the time series of cardiac data and that of wavelets is also important.

The Biorthogonal wavelets are frequently used at the wide field of analysis of images and signals~\cite{chinos}. However, it is not common to hear about the application of a biorthogonal wavelet for the analysis of ECGs. For the analysis of cardiac data, the intrinsic fractality of bior3.1 and bior3.3 (this can be tested graphically, and statistically using Detrended Fluctuation Analysis (DFA)~\cite{physa}) is decisive at scales higher than 64 heartbeats. According to Ref.~\cite{physa}, a time series of interbeat intervals of length $\sim 10^5$ corresponding to healthy people shows fractality considering scales of the order of $10^4$ in DFA and $2^6$ in CVAA. The results in this document indicate that the data collapse obtained using CVAA holds for scales higher than 1024, of the order of $2\times10^4$ heartbeats, which is $\sim 20\%$ of the data. The fitting parameters at so high scales increases slowly as much as 0.3 units. That fractal property of the Biorthogonal wavelets used in our study, suggests the answer to questioning of why data collapse is almost invariant to such high scales discussed here. At the tables~1-4, the fitting parameters corresponding to the wavelet bior3.1, are almost the same up to the highest scale; graphically the data collapse is hold on for the range 64-1024. For other wavelets that situation doesn't happen: As already commented, the fitting parameters change numerically and graphically the data collapsing, and is just hold on functionally, because of the data collapsing is missed dramatically for scales higher than 64. 

The long term and long scale study of the ECG of healthy humans applying the CVAA: The biorthogonal wavelet bior3.1, suggests that does exist an  intrinsic behavior of the cardiac dynamics, that is still universaler than the already discovered at Ref.~\cite{nature}. According to this result, the intrinsic dynamics of a healthy heart isn't affected by any of the possible non stationarities produced during normal activities of a person during a complete day, including the perturbation of the circadian rhythm, letting us to unveil a deeper dynamics at the heart, that is commonly masked by a set of noises of a different kind. The result described here, generalizes that found at Ref.~\cite{nature}, but contradicts it in the sense that there is a wavelet for which the day and night phases under CVAA are not important for the dynamics of the heart. The normalized Gaussian distribution -that characterizes such a behavior- is the solution to the Fokker-Planck (FP) equation, that describes the probability of finding a particle in a fluid, that moves following a dynamics represented by the Langevin equation. The FP equation represents classically a system out of equilibrium. Very recently, it has been found, that the solution to the FP is able to be parametrized according to the Statistics of Tsallis~\cite{tsallis}. According to those studies~\cite{polacos}, the Gamma distribution is :  $G_q(x)=b^{[\nu(q)+1]}x^\nu(q)e^{-b(q)x}/\Gamma[\nu(q)+1]$, where $\nu(q)=(q-1)^{-1}$ being $1<q<2$. It is known that when $q\to 1$, the system is most in equilibrium while when $q\to 2$, the system isn't. Taking into account the results of the para metrical study at Ref.~\cite{polacos}, substituting $\nu=1.43\pm 0.02$ in the last equality, it is found that $q=1.7\pm 0.02$. That value of the non extensive parameter corresponding to the heartbeating of healthy patients, indicates that the heart behaves like a system out of equilibrium --like those represented by the FP equation--, and that kind of nonlinear phenomena is able to be represented by the Non extensive Statistical Mechanics of Tsallis.  

\section{Conclusions.}

Using the CVAA: Wavelet bior3.1 it is found, that the intrinsically collective behavior of the heartbeating of healthy people can be represented according to the Statistical Mechanics of Tsallis, for which corresponds a non extensive parameter $q=1.7\pm 0.02$.

\section*{Acknowledgments.}

The author thanks to twice organisms: Promep for academical support, and to Conacyt, because of the Project of my authorship: Sep-2003-C02-45007. Part of this work was done at Unacar, M\ez xico.

\setlength{\tabcolsep}{4.5pt}

\begin{sidewaystable}
\caption{After the CVAA, these are the parameters that correspond to the fit of the Gamma function to the distribution of amplitudes. The scales shown here are $(1-4)\times2^6$ heartbeats.}

\resizebox{19.5cm}{5cm}{
\bt{|c|c|ccc|ccc|ccc|}
\hline
 & & \mtc{9}{c|}{W\,\,A\,\,V\,\,E\,\,L\,\,E\,\,T} \\
\cline{3-11} 
$n\times2^6$ & & \mtc{3}{c|}{Daubechies} & \mtc{3}{|c|}{Biorthogonal} & \mtc{3}{c|}{Gaussian} \\ 
\cline{3-11}
      & {\begin{rotate}{90}{Params.}\end{rotate}} & db1 & db2 & db3 & bior3.1 & bior3.3 & bior3.5 & gaus1 & gaus2 & gaus3 \\
\hline
\hline

& $  \nu $ & $ 1.18  \pm 0.04 $ & $ 1.16  \pm 0.04 $ & $ 1.18  \pm 0.04 $ & $ 1.41  \pm 0.03 $ & $ 1.24  \pm 0.03 $ & $ 1.22  \pm 0.03 $ & $ 1.08  \pm 0.06 $ & $ 0.96  \pm 0.05 $ & $ 0.99  \pm 0.04 $ \\
1 & $ b $ & $ 3.04  \pm 0.07 $ & $ 3.02  \pm 0.07 $ & $ 3.03  \pm 0.07 $ & $ 3.24  \pm 0.05 $ & $ 3.08  \pm 0.05 $ & $ 3.07  \pm 0.06 $ & $ 2.95  \pm 0.12 $ & $ 2.78  \pm 0.10 $ & $ 2.80  \pm 0.09 $ \\ 
& $ \chi^2 $ & 0.36 & 0.35 & 0.37 & 0.22 & 0.27 & 0.32 & 0.65 & 0.57 & 0.53
\\ \hline
& $  \nu $ & $ 1.06  \pm 0.05 $ & $ 0.97  \pm 0.05 $ & $ 0.99  \pm 0.05 $ & $ 1.41  \pm 0.03 $ & $ 1.18  \pm 0.03 $ & $ 1.10  \pm 0.04 $ & $ 1.13  \pm 0.08 $ & $ 1.00  \pm 0.06 $ & $ 1.01  \pm 0.06 $ \\
2 & $ b $ & $ 2.92  \pm 0.09 $ & $ 2.78  \pm 0.09 $ & $ 2.78  \pm 0.10 $ & $ 3.22  \pm 0.06 $ & $ 3.00  \pm 0.06 $ & $ 2.91  \pm 0.08 $ & $ 2.99  \pm 0.14 $ & $ 2.85  \pm 0.12 $ & $ 2.84  \pm 0.11 $ \\ 
& $ \chi^2 $ & 0.51 & 0.57 & 0.61 & 0.25 & 0.33 & 0.46 & 0.78 & 0.74 & 0.66
\\ \hline
& $  \nu $ & $ 1.08  \pm 0.06 $ & $ 0.94  \pm 0.05 $ & $ 0.97  \pm 0.06 $ & $ 1.37  \pm 0.03 $ & $ 1.07  \pm 0.03 $ & $ 0.99  \pm 0.04 $ & $ 1.11  \pm 0.07 $ & $ 1.08  \pm 0.08 $ & $ 1.10  \pm 0.08 $ \\
3 & $ b $ & $ 2.94  \pm 0.11 $ & $ 2.77  \pm 0.10 $ & $ 2.77  \pm 0.11 $ & $ 3.16  \pm 0.05 $ & $ 2.86  \pm 0.06 $ & $ 2.77  \pm 0.08 $ & $ 2.91  \pm 0.13 $ & $ 2.90  \pm 0.15 $ & $ 2.93  \pm 0.14 $ \\ 
& $ \chi^2 $ & 0.62 & 0.63 & 0.68 & 0.25 & 0.34 & 0.51 & 0.76 & 0.89 & 0.83\\ \hline
& $  \nu $ & $ 1.11  \pm 0.07 $ & $ 0.96  \pm 0.06 $ & $ 0.98  \pm 0.06 $ & $ 1.40  \pm 0.03 $ & $ 1.12  \pm 0.04 $ & $ 1.07  \pm 0.06 $ & $ 1.15  \pm 0.07 $ & $ 1.02  \pm 0.07 $ & $ 1.07  \pm 0.06 $ \\
4 & $ b $ & $ 2.98  \pm 0.13 $ & $ 2.78  \pm 0.11 $ & $ 2.77  \pm 0.12 $ & $ 3.19  \pm 0.05 $ & $ 2.92  \pm 0.08 $ & $ 2.88  \pm 0.11 $ & $ 2.93  \pm 0.13 $ & $ 2.81  \pm 0.13 $ & $ 2.84  \pm 0.12 $ \\ 
& $ \chi^2 $ & 0.69 & 0.70 & 0.77 & 0.50 & 0.70 & 0.66 & 0.83 & 0.83 & 0.80\\
\hline
\et
}
\end{sidewaystable}

\begin{sidewaystable}

\caption{After the CVAA, these are the parameters that correspond to the fit of the Gamma function to the distribution of amplitudes. The scales shown here are $(5-8)\times2^6$ heartbeats.}
\resizebox{19.5cm}{5cm}{
\bt{|c|c|ccc|ccc|ccc|}
\hline
 & & \mtc{9}{c|}{W\,\,A\,\,V\,\,E\,\,L\,\,E\,\,T} \\
\cline{3-11} 
$n\times2^6$ &  & \mtc{3}{c|}{Daubechies} & \mtc{3}{|c|}{Biorthogonal} & \mtc{3}{c|}{Gaussian} \\ 
\cline{3-11}
      & {\begin{rotate}{90}{Params.}\end{rotate}}  & db1 & db2 & db3 & bior3.1 & bior3.3 & bior3.5 & gaus1 & gaus2 & gaus3 \\
\hline
\hline

& $  \nu $ & $ 1.13  \pm 0.07 $ & $ 0.98  \pm 0.06 $ & $ 1.06  \pm 0.07 $ & $ 1.39  \pm 0.03 $ & $ 1.09  \pm 0.04 $ & $ 1.05  \pm 0.06 $ & $ 1.20  \pm 0.08 $ & $ 1.02  \pm 0.07 $ & $ 1.07  \pm 0.06 $ \\
5 & $ b $ & $ 2.99  \pm 0.14 $ & $ 2.81  \pm 0.12 $ & $ 2.86  \pm 0.13 $ & $ 3.18  \pm 0.05 $ & $ 2.89  \pm 0.08 $ & $ 2.83  \pm 0.12 $ & $ 2.96  \pm 0.15 $ & $ 2.78  \pm 0.13 $ & $ 2.86  \pm 0.11 $ \\ 
& $ \chi^2 $ & 0.75 & 0.77 & 0.83 & 0.25 & 0.51 & 0.75 & 0.93 & 0.87 & 0.74
\\ \hline
& $  \nu $ & $ 1.13  \pm 0.07 $ & $ 1.10  \pm 0.09 $ & $ 1.14  \pm 0.09 $ & $ 1.39  \pm 0.03 $ & $ 1.10  \pm 0.04 $ & $ 1.06  \pm 0.06 $ & $ 1.23  \pm 0.09 $ & $ 1.08  \pm 0.07 $ & $ 1.05  \pm 0.07 $ \\
6 & $ b $ & $ 2.98  \pm 0.14 $ & $ 2.93  \pm 0.16 $ & $ 2.94  \pm 0.16 $ & $ 3.17  \pm 0.04 $ & $ 2.90  \pm 0.08 $ & $ 2.84  \pm 0.11 $ & $ 2.98  \pm 0.17 $ & $ 2.85  \pm 0.13 $ & $ 2.81  \pm 0.12 $ \\ 
& $ \chi^2 $ & 0.79 & 0.94 & 0.98 & 0.22 & 0.49 & 0.72 & 1.06 & 0.87 & 0.81
\\ \hline
& $  \nu $ & $ 1.10  \pm 0.08 $ & $ 1.09  \pm 0.09 $ & $ 1.16  \pm 0.10 $ & $ 1.43  \pm 0.03 $ & $ 1.12  \pm 0.05 $ & $ 1.14  \pm 0.09 $ & $ 1.24  \pm 0.09 $ & $ 1.13  \pm 0.08 $ & $ 1.13  \pm 0.07 $ \\
7 & $ b $ & $ 2.93  \pm 0.14 $ & $ 2.90  \pm 0.16 $ & $ 2.94  \pm 0.17 $ & $ 3.20  \pm 0.04 $ & $ 2.91  \pm 0.08 $ & $ 2.93  \pm 0.15 $ & $ 2.99  \pm 0.16 $ & $ 2.90  \pm 0.14 $ & $ 2.89  \pm 0.13 $ \\ 
& $ \chi^2 $ & 0.82 & 1.00 & 1.08 & 0.24 & 0.51 & 0.94 & 1.01 & 0.95 & 0.85
\\ \hline
& $  \nu $ & $ 1.09  \pm 0.07 $ & $ 1.07  \pm 0.09 $ & $ 1.16  \pm 0.11 $ & $ 1.43  \pm 0.03 $ & $ 1.12  \pm 0.05 $ & $ 1.10  \pm 0.07 $ & $ 1.33  \pm 0.10 $ & $ 1.13  \pm 0.08 $ & $ 1.19  \pm 0.08 $ \\
8 & $ b $ & $ 2.90  \pm 0.14 $ & $ 2.85  \pm 0.17 $ & $ 2.91  \pm 0.19 $ & $ 3.20  \pm 0.04 $ & $ 2.90  \pm 0.09 $ & $ 2.85  \pm 0.14 $ & $ 3.07  \pm 0.17 $ & $ 2.88  \pm 0.15 $ & $ 2.97  \pm 0.14 $ \\ 
& $ \chi^2 $ & 0.84 & 1.05 & 1.19 & 0.46 & 0.54 & 0.91 & 1.05 & 1.05 & 0.90
\\ \hline
\et
}
\end{sidewaystable}

\begin{sidewaystable}

\caption{After the CVAA, these are the parameters that correspond to the fit of the Gamma function to the distribution of amplitudes. The scales shown here are $(9-12)\times2^6$ heartbeats.}
\resizebox{19.5cm}{5cm}{
\bt{|c|c|ccc|ccc|ccc|}
\hline
 & & \mtc{9}{c|}{W\,\,A\,\,V\,\,E\,\,L\,\,E\,\,T} \\
\cline{3-11} 
$n\times2^6$ & & \mtc{3}{c|}{Daubechies} & \mtc{3}{|c|}{Biorthogonal} & \mtc{3}{c|}{Gaussian} \\ 
\cline{3-11}
      & {\begin{rotate}{90}{Params.}\end{rotate}} & db1 & db2 & db3 & bior3.1 & bior3.3 & bior3.5 & gaus1 & gaus2 & gaus3 \\
\hline
\hline

& $  \nu $ & $ 1.11  \pm 0.07 $ & $ 1.05  \pm 0.09 $ & $ 1.18  \pm 0.12 $ & $ 1.44  \pm 0.03 $ & $ 1.10  \pm 0.05 $ & $ 1.09  \pm 0.08 $ & $ 1.44  \pm 0.11 $ & $ 1.12  \pm 0.08 $ & $ 1.22  \pm 0.08 $ \\
9 & $ b $ & $ 2.90  \pm 0.14 $ & $ 2.83  \pm 0.17 $ & $ 2.94  \pm 0.20 $ & $ 3.20  \pm 0.04 $ & $ 2.85  \pm 0.09 $ & $ 2.82  \pm 0.14 $ & $ 3.17  \pm 0.17 $ & $ 2.85  \pm 0.14 $ & $ 2.99  \pm 0.14 $ \\ 
& $ \chi^2 $ & 0.82 & 1.06 & 1.22 & 0.23 & 0.57 & 0.99 & 1.05 & 0.99 & 0.95
\\ \hline
& $  \nu $ & $ 1.10  \pm 0.07 $ & $ 1.04  \pm 0.09 $ & $ 1.16  \pm 0.12 $ & $ 1.45  \pm 0.03 $ & $ 1.12  \pm 0.05 $ & $ 1.10  \pm 0.08 $ & $ 1.53  \pm 0.11 $ & $ 1.20  \pm 0.08 $ & $ 1.25  \pm 0.08 $ \\
10 & $ b $ & $ 2.88  \pm 0.14 $ & $ 2.80  \pm 0.17 $ & $ 2.92  \pm 0.20 $ & $ 3.21  \pm 0.04 $ & $ 2.87  \pm 0.09 $ & $ 2.83  \pm 0.14 $ & $ 3.24  \pm 0.18 $ & $ 2.91  \pm 0.15 $ & $ 3.00  \pm 0.15 $ \\ & $ \chi^2 $ & 0.82 & 1.13 & 1.22 & 0.23 & 0.62 & 0.97 & 1.01 & 1.07 & 0.98
\\ \hline
& $  \nu $ & $ 1.12  \pm 0.08 $ & $ 0.98  \pm 0.09 $ & $ 1.06  \pm 0.10 $ & $ 1.45  \pm 0.03 $ & $ 1.12  \pm 0.05 $ & $ 1.12  \pm 0.09 $ & $ 1.68  \pm 0.12 $ & $ 1.23  \pm 0.08 $ & $ 1.22  \pm 0.08 $ \\
11 & $ b $ & $ 2.89  \pm 0.14 $ & $ 2.73  \pm 0.16 $ & $ 2.81  \pm 0.19 $ & $ 3.21  \pm 0.04 $ & $ 2.89  \pm 0.09 $ & $ 2.87  \pm 0.16 $ & $ 3.40  \pm 0.19 $ & $ 2.95  \pm 0.14 $ & $ 2.97  \pm 0.14 $ \\ 
& $ \chi^2 $ & 0.85 & 1.11 & 1.24 & 0.23 & 0.63 & 1.07 & 1.03 & 1.06 & 0.95
\\ \hline
& $  \nu $ & $ 1.13  \pm 0.08 $ & $ 0.98  \pm 0.09 $ & $ 1.04  \pm 0.10 $ & $ 1.43  \pm 0.03 $ & $ 1.15  \pm 0.06 $ & $ 1.14  \pm 0.09 $ & $ 1.77  \pm 0.13 $ & $ 1.29  \pm 0.09 $ & $ 1.26  \pm 0.09 $ \\
12 & $ b $ & $ 2.90  \pm 0.14 $ & $ 2.71  \pm 0.17 $ & $ 2.76  \pm 0.18 $ & $ 3.19  \pm 0.04 $ & $ 2.93  \pm 0.11 $ & $ 2.92  \pm 0.16 $ & $ 3.48  \pm 0.20 $ & $ 3.01  \pm 0.15 $ & $ 3.01  \pm 0.15 $ \\ & $ \chi^2 $ & 0.87 & 1.18 & 1.27 & 0.24 & 0.69 & 1.04 & 1.05 & 1.11 & 1.03\\ 
\hline
\et
}
\end{sidewaystable}

\begin{sidewaystable}

\caption{After the CVAA, these are the parameters that correspond to the fit of the Gamma function to the distribution of amplitudes. The scales shown here are $(13-16)\times2^6$ heartbeats.}
\resizebox{19.5cm}{5cm}{
\bt{|c|c|ccc|ccc|ccc|}
\hline
 & & \mtc{9}{c|}{W\,\,A\,\,V\,\,E\,\,L\,\,E\,\,T} \\
\cline{3-11} 
$n\times2^6$ & & \mtc{3}{c|}{Daubechies} & \mtc{3}{|c|}{Biorthogonal} & \mtc{3}{c|}{Gaussian} \\ 
\cline{3-11}
      &{\begin{rotate}{90}{Params.}\end{rotate}}& db1 & db2 & db3 & bior3.1 & bior3.3 & bior3.5 & gaus1 & gaus2 & gaus3 \\
\hline
\hline

& $  \nu $ & $ 1.15  \pm 0.09 $ & $ 1.07  \pm 0.11 $ & $ 1.12  \pm 0.11 $ & $ 1.47  \pm 0.03 $ & $ 1.18  \pm 0.06 $ & $ 1.10  \pm 0.08 $ & $ 1.81  \pm 0.13 $ & $ 1.34  \pm 0.09 $ & $ 1.29  \pm 0.09 $ \\
13 & $ b $ & $ 2.92  \pm 0.16 $ & $ 2.79  \pm 0.19 $ & $ 2.85  \pm 0.20 $ & $ 3.23  \pm 0.05 $ & $ 2.96  \pm 0.11 $ & $ 2.88  \pm 0.15 $ & $ 3.51  \pm 0.19 $ & $ 3.04  \pm 0.16 $ & $ 3.01  \pm 0.15 $ \\ & $ \chi^2 $ & 0.93 & 1.29 & 1.31 & 0.26 & 0.70 & 0.98 & 1.03 & 1.12 & 1.01\\ 
\hline
& $  \nu $ & $ 1.16  \pm 0.09 $ & $ 1.07  \pm 0.10 $ & $ 1.17  \pm 0.11 $ & $ 1.45  \pm 0.03 $ & $ 1.09  \pm 0.05 $ & $ 1.04  \pm 0.09 $ & $ 1.86  \pm 0.13 $ & $ 1.40  \pm 0.10 $ & $ 1.28  \pm 0.09 $ \\
14 & $ b $ & $ 2.93  \pm 0.16 $ & $ 2.80  \pm 0.17 $ & $ 2.88  \pm 0.20 $ & $ 3.21  \pm 0.04 $ & $ 2.85  \pm 0.10 $ & $ 2.76  \pm 0.16 $ & $ 3.55  \pm 0.20 $ & $ 3.10  \pm 0.16 $ & $ 3.00  \pm 0.15 $ \\ & $ \chi^2 $ & 0.93 & 1.19 & 1.32 & 0.24 & 0.66 & 1.09 & 1.06 & 1.16 & 1.06\\ 
\hline
& $  \nu $ & $ 1.18  \pm 0.09 $ & $ 1.13  \pm 0.11 $ & $ 1.26  \pm 0.13 $ & $ 1.46  \pm 0.03 $ & $ 1.08  \pm 0.06 $ & $ 1.08  \pm 0.10 $ & $ 1.78  \pm 0.12 $ & $ 1.46  \pm 0.11 $ & $ 1.27  \pm 0.09 $ \\
15 & $ b $ & $ 2.96  \pm 0.17 $ & $ 2.84  \pm 0.19 $ & $ 2.99  \pm 0.22 $ & $ 3.22  \pm 0.05 $ & $ 2.81  \pm 0.10 $ & $ 2.77  \pm 0.17 $ & $ 3.49  \pm 0.19 $ & $ 3.15  \pm 0.17 $ & $ 2.95  \pm 0.16 $ \\ & $ \chi^2 $ & 0.97 & 1.28 & 1.42 & 0.25 & 0.69 & 1.21 & 1.02 & 1.18 & 1.13\\
\hline
& $  \nu $ & $ 1.17  \pm 0.09 $ & $ 1.14  \pm 0.12 $ & $ 1.31  \pm 0.14 $ & $ 1.46  \pm 0.03 $ & $ 1.09  \pm 0.06 $ & $ 1.19  \pm 0.10 $ & $ 1.81  \pm 0.13 $ & $ 1.64  \pm 0.12 $ & $ 1.37  \pm 0.11 $ \\
16 & $ b $ & $ 2.94  \pm 0.17 $ & $ 2.87  \pm 0.20 $ & $ 3.06  \pm 0.24 $ & $ 3.23  \pm 0.05 $ & $ 2.83  \pm 0.11 $ & $ 2.91  \pm 0.18 $ & $ 3.52  \pm 0.20 $ & $ 3.31  \pm 0.19 $ & $ 3.07  \pm 0.18 $ \\ 
& $ \chi^2 $ & 1.00 & 1.34 & 1.49 & 0.52 & 0.75 & 1.18 & 1.06 & 1.22 & 1.21\\
\hline
\et
}
\end{sidewaystable}

\newpage

\bs

\bfi
\epsfig{file=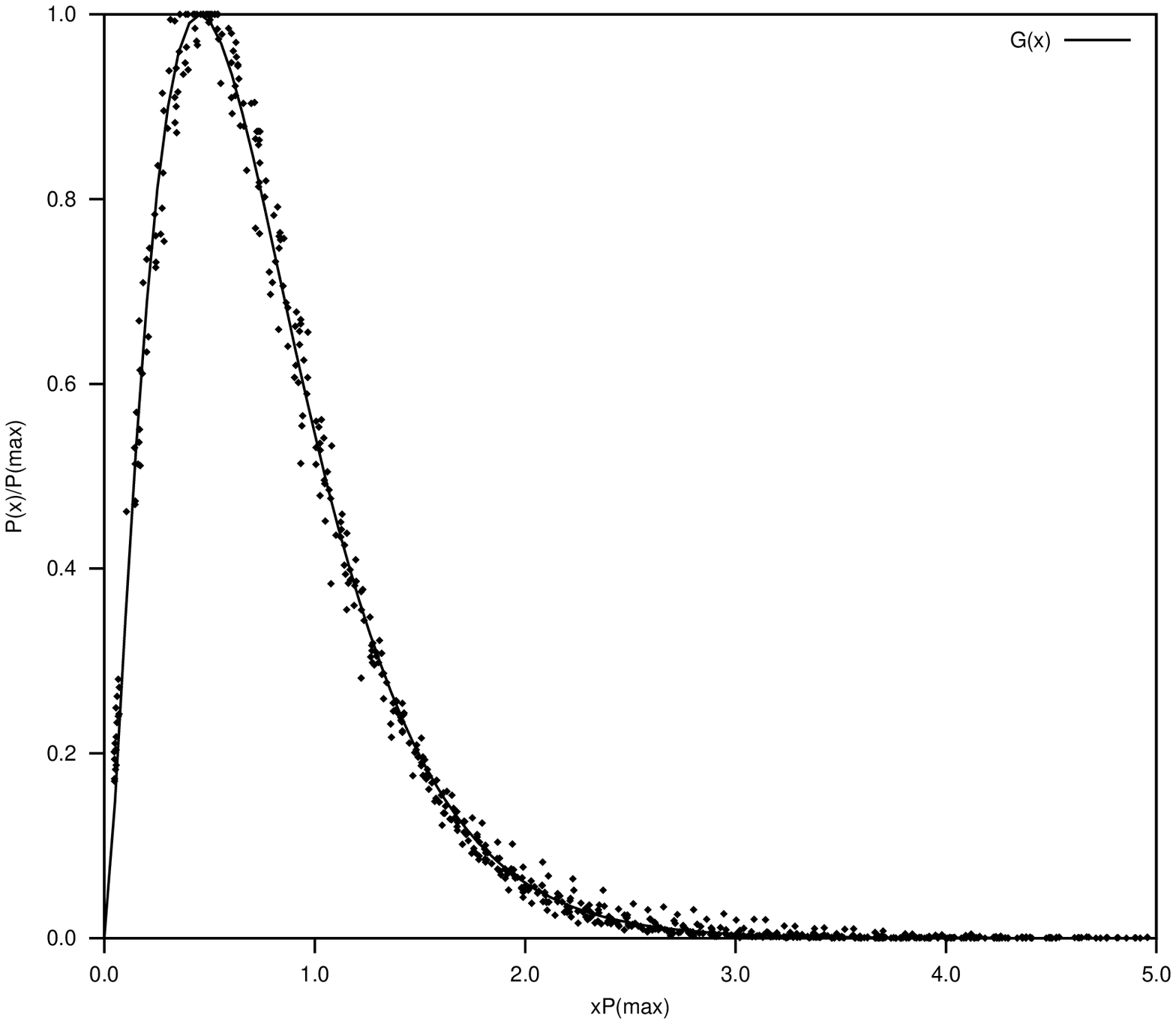,width=1.0\textwidth}
\caption{A distribution of the data (small black squares) corresponding to NSRDB. The parameters of the fit are $\nu=1.46\pm 0.03$ and $\chi^2=0.26$.}
\efi

\end{document}